\documentclass[amssymb, nofootinbib, superscriptaddress]{revtex4}

\usepackage{amsmath}
\usepackage{amsfonts}
\usepackage{graphicx}
\usepackage{psfrag}
\usepackage{array}
\usepackage{bbm}

\newcommand{\f}{\frac}
\newcommand{\tr}{\mathrm{tr}}
\newcommand{\rar}{\rightarrow}
\newcommand{\pd}{\partial}

\newcommand{\su}{\mathfrak{su}}
\newcommand{\SU}{\mathrm{SU}}
\newcommand{\SO}{\mathrm{SO}}
\newcommand{\isu}{\mathfrak{isu}}

\newcommand{\R}{\mathbb{R}}
\newcommand{\C}{\mathbb{C}}
\newcommand{\N}{\mathbb{N}}

\newcommand{\cD}{{\cal D}}
\newcommand{\cK}{{\cal K}}
\newcommand{\cM}{{\cal M}}
\newcommand{\cN}{{\cal N}}
\newcommand{\cO}{{\cal O}}
\newcommand{\cS}{{\cal S}}
\newcommand{\cZ}{{\cal Z}}

\newcommand{\es}{{e^*}}
\newcommand{\vs}{{v^*}}
\newcommand{\fs}{{f^*}}

\newcommand{\id}{\mathbb{I}}

\newcommand{\be}{\begin{equation}}
\newcommand{\ee}{\end{equation}}
\newcommand{\bes}{\begin{eqnarray}}
\newcommand{\ees}{\end{eqnarray}}

\def\tl{\widetilde}
\def\tdelta{\tl{\delta}}
\def\arr{\rightarrow}
\def\eps{\epsilon}
\def\pp{\partial}
\def\ka{\kappa}
\def\cc{{\cal C}}
\def\la{\langle}
\def\ra{\rangle}

\begin{document}
%
\title{\Large\bf A Note on B-observables in Ponzano-Regge 3d Quantum Gravity}


\author{Etera R. Livine}\email{etera.livine@ens-lyon.fr}
\affiliation{Laboratoire de Physique, ENS Lyon, CNRS UMR 5672, 46 All\'ee d'Italie, 69007 Lyon, France}

\author{James P. Ryan}\email{jryan@perimeterinstitute.ca}
\affiliation{Perimeter Institute, 31 Caroline St. N., Waterloo, ON N2L 2Y5, Canada}

\date{\small \today}


\begin{abstract}
We study the insertion and value of metric observables in the (discrete) path integral formulation
of the Ponzano-Regge spinfoam model for 3d quantum gravity. In particular, we discuss the length
spectrum and the relation between insertion of such B-observables and gauge fixing in the path
integral.
\end{abstract}

\maketitle

\section{Introduction}
\label{intro}

As it stands, it is technically extremely difficult to extract physical information from gravitational theories, both classical and quantum level.  In both regimes, this process requires the definition of observables, which should be both invariant under the symmetries of the theory, and nonetheless, physically relevant.  Such observables are hard to come by, let alone compute.    At the classical level, the majority of work has been done within the canonical setting.   Only recently, has the road been opened to fully calculate observables within this formalism \cite{bd}, using the technique of partial observables \cite{cr}.

At the quantum level, however, almost nothing is known of physical geometric observables.  Some progress has been made within both the canonical quantisation process and the path integral approach.  In loop quantum gravity \cite{lqg}, the geometric observables lie mostly in the domain of the kinematical Hilbert space where the volume and area operators are defined and are diagonal on the space of states \cite{geometric}.  In the realm of covariant approaches to quantum gravity, a systematic analysis of the space of physical geometric observables has remained untouched.   It is to redress this balance that we apply our resources hereafter.

On the other hand, much progress has been made recently, within both the canonical and covariant formalisms, to evaluate matter observables \cite{classical, quantumc, PR1, quantump} (both at the classical and quantum level).  Within the discrete path integral approach known as spin foams \cite{baez, principle}, the effective matter observable have been shown to arise as the Feynman diagrams of  a non-commutative quantum field theory \cite{PR3, effqg}. Interestingly, matter has also been dealt with in the regime of group field theory \cite{gft}; a theory which generates spin foams as Feynman diagrams.  Recently, the non-commutative field theory encountered above has been shown to arise as a particular perturbation around a classical solution of the group field theory \cite{phase}.

We shall also follow the route of spin foams to define our path integral.  These are sums over configurations with support on piecewise linear manifolds.  The amplitudes are purely combinatorial and depend on the representation theory of the appropriate gauge group.  They may be recast in a form which resembles the path integral of discretised (constrained) $BF$ theories.   It is in this form, that we will perform our calculations.  In the case of 3d gravity, there are no constraints, and we are left with the simple $BF$ term.  Therefore, the possible observables are gauge invariant functions of the $B$ variables and the holomonies.

We shall pay most attention here to geometric variables such as the length squared, and  the dihedral angle between edges.   To get meaningful results, one must gauge-fix the amplitude and we pay special attention here to this process.  In the usual scenario \cite{lfdl}, gauge fixing is the process of picking maximal tree with the relevant discrete structures and trivialising the variables on those edges of the trees.  In general, a Fadeev-Popov determinant will arise, which we must take into account.  We invoke this process here, also.
To extract more information from the observables,  we introduce normalised \lq\lq states'' which serve to pick out the edge of the triangulation upon which we intend to evaluate the observable.

Another motivation, is that $B$ functionals, occur frequently, when we rewrite theories, such as Yang-Mills and 4d quantum gravity, in terms of constrained $BF$ theories \cite{principle, pleb}.  Thus, understanding their nature in this simplified setting will pave the way to dealing with them more generally.

The outline of the paper is a follows.  In Section \ref{qg}, we outline the passage from the classical continuum theory, to the discrete path integral approach.  In succeeding subsections, we recover the $\SU(2)$ Ponzano-Regge model from the $\SO(3)$ theory, and discuss the ambiguity in the discretisation of the $B$ field.  We then proceed to evaluate the expectation value of $B$ observables in Section \ref{bins}.  We reserve a more thorough investigation of the gauge-fixing procedure for Section \ref{observables}, and subsequently present our conclusions and outlook.

\section{Discrete path integral for the Ponzano-Regge model}
\label{qg}

Before we commence to describe the main import of our work,  we shall lay out succinctly the particular quantum gravity theory which provides the framework within
which we shall calculate.   For our purposes, it is sufficient to restrict ourselves to a 3d
Riemannian theory of gravity without cosmological constant.  To pass to a quantum theory, we shall
write the action in $1st$ order form as an $\su(2)$ gauge theory:
\be
\cS[B,\omega] = \int_{\cM}\tr(B\wedge F[\omega]).
\ee
$\cM$ is a closed manifold. The triad $B$ is a $\su(2)$-valued 1-form, from which the
metric is reconstructed as $g_{\mu\nu} \equiv B_\mu^iB_\nu^j\eta_{ij}$. The parallel transport on
the manifold is given by the $\su(2)$-valued connection $\omega$ and its curvature $F[\omega] =
d\omega +\omega\wedge\omega$ is an $\su(2)$-valued 2-form. $\tr$ is the trace over the Lie algebra.
The classical equations of motion are:
\be
d_{\omega}B \equiv dB +  [\omega, B] = 0, \quad\quad F[w]  = 0
\ee
where $[\cdot,\cdot]$ is the Lie bracket on the algebra. These equations impose that the connection
is torsion-free and flat.  Solving for the connection in terms of the triad  returns us to the
standard $2nd$ order metric theory for 3d Riemannian gravity. $i,j$ are $\su(2)$ indices and
$\eta_{ij}$ is the flat metric on $\su(2)$, while $\mu,\nu$ are \lq space-time' indices.

This action has several symmetries; namely, it is invariant under rotations, translations and
diffeomorphisms.\footnote{At the continuum level, we write down the gauge symmetries of this action:
 \bes
 \label{}
 \textrm{Rotation}
 &\left\{
 \begin{array}{l}
 \omega  \rar  k^{-1}dk + k^{-1}\omega k\\
 B  \rar k^{-1}Bk
 \end{array}\right.
 & \;\textrm{parametrised by $k\in\SU(2)$}, \\
 \textrm{Translation}
 &\left\{
 \begin{array}{l}
 \omega \rar \omega \\
 B \rar B + d_{\omega}\phi
 \end{array}\right.\hphantom{xxxxx}\,
 & \; \textrm{parametrised by $\phi \in \su(2)$}.
\ees
The translation symmetry holds due to the Bianchi identity, $d_\omega F[\omega] \equiv 0$, provided
$\phi=0$ on the boundaries $\pd\cM$ (i.e. the translation symmetry does not extend to the
boundary).  Remarkably, infinitesimal $\isu(2)$ symmetry is equivalent {\it on-shell} to
diffeomorphism symmetry provided $\det(B)\neq0$.  We see this by the following table of
transformations:
\be
\label{}
\begin{array}{ll}
\delta^R_{\rho}\, \omega = d_\omega\rho\, &    \delta^R_{\rho}\, B = [B, \rho],\\
\delta^T_{\phi}\, \omega = 0, &    \delta^T_{\phi} \, B = d_\omega\phi,\\
\delta^D_{\xi}\, \omega = \delta^R_{\imath_{\xi}\omega}\, \omega + \delta^T_{\imath_{\xi}B}\, \omega + \imath_{\xi}(F[\omega]), &
\delta^D_{\xi}\, B = \delta^R_{\imath_{\xi}\omega}\, B + \delta^T_{\imath_{\xi}B}\, B + \imath_{\xi}(d_\omega B),
\end{array}
\ee
where the diffeomorphism, parametrised by a vector $\xi$, acts like a Lie derivative:
$\delta^D_{\xi} \equiv d\imath_{\xi} + \imath_\xi d$.\footnote{$d$ is the exterior product on forms
and $\imath_\xi$ is the interior product.}  The symmetry becomes manifest on-shell if we choose
$\rho = \imath_\xi \omega$ and $\phi = \imath_\xi B$.} These symmetries require proper
gauge-fixing. Nevertheless, we can formally define the partition function\footnote{ This should not be confused with $3d$ Euclidean quantum gravity where the
coefficient of the action in the partition function is $-1$ rather than $i$ and may be regarded as
an analytic continuation of the Lorentzian theory.}:
\be
\cZ_{\cM} = \int \cD B\, \cD\omega\;  e^{i\cS[B,\omega]}.
\ee
The triad field $B$ is interpreted as a Lagrange multiplier enforcing the flatness of the
connection: $\cZ=\int \cD\omega \delta(F[\omega]))$. As the $BF$ theory action is topological, we
may apply methods developed in topological quantum field theory to evaluate this partition function
and appropriate gauge-invariant observables. This is not our aim here; we wish to study the link
with discrete models of quantum gravity and compute the spectrum of geometrical operators
(observables depending on the $B$-field) in a discrete path integral formalism.

With this target in mind, we replace the manifold $\cM$ by a simplicial manifold
$\Delta$ of the same topology. Heuristically speaking, since the $BF$ theory is topological and
does not have any local degree of freedom, we do not expect to lose information in this
replacement. In three space-time dimensions, we can triangulate any manifold. We label the 0-,1-,2-
and 3-subsimplices as $v,e,f$ and $t$, respectively.  Another important constituent is the
topological dual $\Delta^*$ to the simplicial complex.  We label sub-elements of this structure as
$v^*,e^*,f^*$ and $t^*$, respectively.  The fields $B$ and $w$ are replaced by configurations which
are distributional with support on subsimplices of $\Delta$ and its topological dual $\Delta^*$. We
integrate these fields over the appropriate subsimplices. The definition of the integrated fields
is:
\be\setlength{\extrarowheight}{0.2cm}
\begin{array}{lcll}
B           & \rar &    B_e         = \int_eB                   & \in \su(2),\\
\omega      & \rar &    g_{e^*}     = {\cal P}e^{\int_{e^*}\omega}      &  \in \SU(2),\\
F[\omega]   & \rar &    G_e         = \prod_{e^*\in \pd f^*} g_{e^*}    & \in \SU(2).
\end{array}
\ee
The flatness constraint $F[\omega]=0$ translates into the triviality of holonomies  $G_e=\id$
(around closed loops, for trivial homotopy). At the discrete level, we will therefore replace the
$\delta(F[\omega])$ constraint by $\delta(G_e)$ constraints, with the discrete $B_e$ variables
still playing the role of the Lagrange multipliers. Following \cite{PR1}, the action on the
simplicial manifold then reads:
\be
\cS[B_e, g_{\es}] = \sum_{e\in\Delta} \tr(B_eG_e),
\ee
where the precise definition of the trace will be given later. Comparing this discrete action with
the continuous BF action, we have simply replaced the curvature $F$ by the holonomy $G$. The first
remark concerns the definition of the holonomy $G_e$ as the oriented product of the group elements
$g_{e^*}$ for all dual edges $e^*$ around the dual face (or plaquette) $f^*$ corresponding to the
considered edge $e$. In order to define such a holonomy, one needs to choose a starting point along
the loop. Changing the starting point amounts to acting on $G_e$ by conjugation by some $\SU(2)$
group element $k$. This is compensated by rotating the $B$-variable, $B_e\arr k B_e k^{-1}$.
Therefore, as long as we restrict ourselves to considering gauge invariant observables of the $B$
variables, such as the norm of $B_e$,  there is no issue. The second remark is about the difference
between $F[\omega]$ which lives in the Lie algebra $\su(2)$ and the holonomy $G_e$ defined as a
group element in $\SU(2)$. Actually the original path integral derivation of the Ponzano-Regge
model introduced the Lie algebra element $Z_e = \log(G_e)\in
\su(2)$ and considered the action $\sum_ e \tr(B_e Z_e)$  \cite{principle}. Nevertheless, one faces
the issue of the non-continuity of the $\log$ map (and the choice of a particular branch and so
on). Therefore, we prefer to work with the formulation presented in \cite{PR1} which uses directly
the group element $G_e$.

The continuous gauge symmetries are broken, but there are residual discrete gauge symmetries: the
rotation symmetry acts at the vertices $\vs\in \Delta^*$ while the translation symmetry acts at
$v\in\Delta$.\footnote{The gauge symmetries are:
\bes
 \textrm{Rotation}  &\left\{
        \begin{array}{l}
        g_{\es}  \rar k_{t(\es)}^{-1}g_{\es}k_{s(\es)}\;\phantom{xxxxxxxxxxxxxx}\\
        B_e  \rar k_{\vs(e)}^{-1}B_ek_{\vs(e)}
        \end{array}\right.
    \phantom{xxxxxxxxxxxxxxxxxxxxxxxx}  \textrm{parametrised by $k_{\vs}\in\SU(2)$},\;\quad \\
 \textrm{Translation}   &\left\{
        \begin{array}{l}
        g_{\es} \rar g_{\es} \\
        B_e \rar B_e + U_e^{t(e)}\phi_{t(e)} - [\Omega^{t(e)}_e, \phi_{t(e)}] -  U_e^{s(e)}\phi_{s(e)} + [\Omega^{s(e)}_e, \phi_{s(e)}] \label{disctrans}
        \end{array}\right.
    \phantom{xxxxx} \textrm{parametrised by $\phi_v \in \su(2)$},\quad\quad
\ees
where $s(\es),\, t(\es)$ are the source and target vertices of $\es$, and $\vs(e)$ is the vertex
where the holonomy around the face $\fs$ begins.  It does not matter which vertex we choose as we
may transfer to any vertex $\vs\subset\pd\fs$ by conjugation. Finally, $\Omega^v_e = \int_e \omega$
starting from $v$, in the limit where the edgelength is small \cite{PR1}, and $U_e^{v}$ is a precise function of the holonomies given in Appendix \ref{fpdet}. }
For details on the action of these discrete gauge symmetries and their gauge fixing, we refer the
interested reader to \cite{PR1}.

Finally, the discrete path integral reads:
\be
\cZ_\Delta = \prod_{e\in\Delta}\int_{\su(2)} d^3B_e \prod_{\es\in\Delta^*}\int_{\SU(2)} dg_\es \; e^{i\cS[B_e, g_\es]},
\ee
where  the Lebesgue measure $d^3B_e$  on $\su(2)\sim\mathbb{R}^3$ and the Haar measure $dg_\es$ on $\SU(2)$ are the natural choice for the discretised path integral measure.
Due to the gauge invariance of the theory, to arrive at sensible results we must fix the gauge,
\cite{PR1}. To do so, we pick maximal trees of edges $T\in\Delta$ and $T^*\in\Delta^*$.\footnotemark
\footnotetext{The maximal tree $T$ touches every vertex $v\in\Delta$ but contains no
loops; likewise for $T^*$.}
One then fixes $B_e = 0$ for all $e\in T$ and $g_\es = \id$ for all $\es\in T^*$.  Mercifully, the
Fadeev-Popov determinant for this gauge evaluates to unity. We will discuss this gauge fixing in
more detail in the Section \ref{observables}.

\subsection{Recovering the Ponzano-Regge model}

Our first task is to evaluate the discrete path integral defined above and show its precise
relation to the Ponzano-Regge model. For this purpose, we introduce the following parametrisation of
$\SU(2)$ group elements in the fundamental two-dimensional representation (spin-$\f12$):
$$
G=\cos\theta\,\id \,+\,i\sin\theta \hat{u}.\vec{\sigma}\,=\,\eps\sqrt{1-p^2}\,\id\,+
\,i\vec{p}.\vec{\sigma},
$$
where the matrices $\sigma_k$, $k=1,2,3$,  are the (Hermitian) Pauli matrices normalised so that
$(\sigma_k)^2=\id$ for all $k$'s and $\tr(\sigma_k\sigma_l)=2\delta_{kl}$.
The parameter $\theta\in[0,2\pi]$ is the class angle labeling the equivalence classes of group
elements under conjugation (the rotation angle is $2\theta$), while $\hat{u}\in
\cS^2$ labels the group elements within each equivalence class (it indicates the rotation axis).
There is an obvious identification between the group elements $g(\theta,\hat{u})$ and
$g(-\theta,-\hat{u})$. We can thus restrict the range of the class angle to $\theta\in[0,\pi]$. The
vector $\vec{p}$ is the projection of the group element $G$ onto the Pauli matrices:
\be
\vec{p}(G)=\,\f{1}{2i}\tr\,G\,\vec{\sigma}.
\ee
This vector has a bounded norm $|\vec{p}|^2\le 1$. The sign $\eps=\pm$ is the sign of $\cos\theta$,
that is $\eps=+$ when $\theta\in[0,\f\pi2]$ and $\eps=-$ when $\theta\in[\f\pi2,\pi]$. Finally, we
also introduce the vector $\vec{u}\,\equiv\,\theta\hat{u}$ which satisfies
$G=\exp(i\vec{u}.\vec{\sigma})$.

We can easily express the normalised Haar measure in term of these variables (see e.g.
\cite{PR1,PR3}):
\be
\int_{\SU(2)} dg
\,=\,
\f{2}{\pi}\int_0^\pi \sin^2\theta\,d\theta\,\int_{\cS^2}\f{d^2\hat{u}}{4\pi}
\,=\,
\sum_{\eps=\pm}\f{1}{2\pi^2}\int_{p\le 1}\f{d^3\vec{p}}{\sqrt{1-p^2}}
\,=\,
\f{1}{2\pi^2}\int_{u\le \pi}\left(\f{\sin u}{u}\right)^2\,d^3\vec{u}\,.
\ee

Finally, we introduce the $\SU(2)$ characters $\chi_j(G)$ for $j\in\N/2$ as the trace of $G$ in the
(spin-$j$) representation of dimension $d_j=(2j+1)$:
$$
\chi_j(G)=\f{\sin d_j\theta}{\sin\theta}=U_{2j}(\cos\theta),
$$
where the functions $U_n$ are the Chebyshev polynomials of the second kind.

We are now equipped to study the discrete path integral introduced above. Following \cite{PR1,PR3},
we get:
\be
\cZ_\Delta=\int \prod_{e}\f{d^3B_e}{4\pi} \int \prod_{\es}dg_\es \; e^{\f12\sum_e \tr(B_eG_e)}
\,=\,
\int \f{d^3B_e}{4\pi} \int dg_\es \; e^{i\sum_e \vec{B}_e\cdot\vec{p}(G_e)}.
\ee
Writing $\vec{p}_e\equiv\vec{p}(G_e)$, we can integrate over the $B_e$ variables and obtain the
product of $\delta^{(3)}(\vec{p}_e)$ distributions. Now, the crucial point of the construction is
the following identity:
\be
\int
\f{d^3B}{4\pi}\,e^{i\vec{B}\cdot\vec{p}(G)}=2\pi^2\delta^{(3)}(\vec{p}(G))=\delta(G)+\delta(-G)\,\equiv\tdelta(G),
\ee
where $\tdelta(.)$ is the distribution localising the group element $G$ on the identity in
$\SO(3)$, i.e it does not distinguish $\id$ and $-\id$ as $\SU(2)$ group elements.
Using the Peter-Weyl theorem, we can decompose $\tdelta(.)$ into $\SU(2)$ representations:
\be
\tdelta(G)=2\sum_{j\in\N}d_j\chi_j(G).
\ee
Then, expanding the $\tdelta(G_e)$ distribution in representations and writing explicitly the group
elements $G_e$ as the product of the $g_\es$ holonomies, one can recover the standard formula for
the Ponzano-Regge model as a product of 6j-symbols attached to each tetrahedron of the
triangulation $\Delta$:
$$
\cZ_{\Delta} =  \sum_{\{j_e\in \N\}}\prod_e (2d_{j_e}) \prod_{t\in\Delta} \{6j\}_t.
$$
We insist that the sum is over integer spin representations $j_e\in\N$ and thus this path integral
formulation gives the $\SO(3)$ Ponzano-Regge model (as shown in \cite{PR1}).

To recover the full $\SU(2)$ Ponzano-Regge model with a sum over both integer and half-integer
representations $j_e\in\N/2$, we introduce a new trick. The goal is to kill the $\delta(-G)$ term
in the $\tdelta(G)$ distribution. This can be achieved by slightly modifying the amplitude in the
path integral:
\be\label{fullpart}
\cZ_\Delta=\int \prod_{e}\f{d^3B_e}{4\pi} \int \prod_{\es}dg_\es \; \prod_e \f12\left(1+\f12\tr G_e\right)e^{\f12\tr(B_eG_e)}
\ee
The new factor vanishes at $G=-\id$ and its value is simply 1 at $G=\id$. Actually, we could use
any other factor with the same property and the present one is the simplest such choice. Then,
using the fact that $\tr(G)=\chi_{1/2}(G)$ by definition, one can check that:~\footnotemark
\be
\f12\left(1+\f12\tr G\right)\,\tdelta(G)
\,=\,\sum_{j\in\N/2} d_j\chi_j(G)
\,=\,\delta(G).
\ee
\footnotetext{A standard formula in the recoupling theory of $\SU(2)$ representations gives for all $j\in\N^*/2$:
$$
\chi_j\chi_{\f12}=\chi_{j+\f12}+\chi_{j-\f12}.
$$
This leads to the following formula for any sequence of coefficients $\{\alpha_j\}_{j\in\N}$:
$$
\left(1+\f12\chi_{\f12}\right)\sum_{j\in\N}\alpha_j \chi_j
\,=\,
\sum_{j\in\N}\alpha_j \chi_j +\sum_{j\in\N+\f12}\f12(\alpha_{j-\f12}+\alpha_{j+\f12}) \,\chi_j.
$$
We also point out the fact that $\chi_j(-G)=(-1)^{2j}\chi_j(G)$.}
At the end of the day, this final path integral with the extra factors reproduces exactly the
Ponzano-Regge amplitudes for 3d quantum gravity including the sum over all odd and even
representations of $\SU(2)$.

\subsection{Generalising the path integral}

Up to now, we have worked with an action $S=\f1{2i}\sum_e \tr (B_eG_e)=\sum_e
\vec{B}_e\cdot\vec{p}(G_e)$. We could however use any other vector $\vec{\phi}(G)$ instead of
$\vec{p}(G)$ as long as the distribution $\delta^{(3)}(\vec{\phi})$ co\"\i ncides with $\delta(G)$
(or $\tdelta(G)$). We also require that $\vec{\phi}$ has a nice behavior under $\SU(2)$
conjugation: if we act on the group element $G\arr k G k^{-1}$ with $k\in\SU(2)$, then the vector
should simply rotate as a 3d vector $\vec{\phi}\arr k\,\vec{\phi}$.

Taking into account these two conditions, we introduce a whole family of possible actions:
$$
S=\sum_e\vec{B}_e\cdot\vec{\phi}(G_e),\qquad
\textrm{with}\quad \vec{\phi}(G)= \phi(\theta)\,\hat{u},
$$
where $\theta\in[0,\pi]\mapsto\phi(\theta)\in\R$ is a continuous map which vanishes only at
$\theta=0$ and possibly also at $\theta=\pi$.

A subclass of actions is given by (continuous) positive functions of the vector $\vec{p}$:
$\vec{\phi}\equiv\,f(p)\,\vec{p}$ with $f(p)\ge 0$ is a rescaling of $\vec{p}$ depending on its
norm. The advantage of this subclass of actions is that $\vec{p}(G)$ is a nicely-behaved polynomial
of the group element $G$. The particularity of this choice is that it is symmetric under
$\theta\arr
\pi-\theta$ and will never distinguish $\id$ from $-\id$. Thus it will naturally lead to the
$\tdelta(G)$ distribution unless we use the extra factor which we discussed above.

Other possibilities shall not be expressed solely in term of $\vec{p}$ and will in general also depend
on the sign $\eps$ (i.e whether $\theta$ is smaller or larger than $\pi/2$). A natural example is
simply $\vec{\phi}=\vec{u}=\theta\,\hat{u}$. Unfortunately it is not a polynomial of the group
element $G$. There are other interesting choices such as
$\vec{\phi}=-i\tr(G\vec{\sigma})/\tr(G)=\tan\theta\,\hat{u}$ but we will not discuss them
further.\footnotemark

\footnotetext{We can introduce the following vector for a group element $G\in\SU(2)$:
\be\label{defc}
\vec{w}=\f{\tr(G\sigma)}{i\tr G}\,=\,\tan\theta\,\hat{u}.
\ee
Contrary to the vectors $\vec{p}$ or $\vec{u}$, the range of $\vec{w}$ is not constrained and
sweeps the whole $\R^3$. The normalised Haar measure reads in term of $\vec{w}$:
\be\label{defm}
\int_{\SU(2)} dg\,=\,
\f{1}{\pi^2}\int_{\R^3}\f{d^3\vec{w}}{(1+w^2)^2}.
\ee
This measure is actually of the type $d^n p_\mu\, /(1+p^2/\ka^2)^\sigma$ usually considered when
deriving generalized uncertainty principle (GUPs) taking into account a minimal length in
phenomenological models for quantum gravity.}

\medskip

Since all these choices define the same path integral and lead to the same partition function (up
to the $\delta(.)$ vs $\tdelta(G)$ subtlety), the difference between these different formulations
resides in the Lagrange multiplier i.e the $B$-field. Indeed, the Lagrange multipliers differ by a
$G$-dependent rescaling which will affect the value of observables depending on the $B$ variables.
More precisely, we write the generalised path integral using an arbitrary $\vec{\phi}$ vector:
$$
\cZ^{(\phi)}_\Delta
\,=\,
\int d^3X_e \int dg_\es \; e^{i\sum_e \vec{X}_e\cdot\vec{\phi}(G_e)},
$$
where we replace the $B_e$ variables by  $X_e$ variables to underline that they do not represent
the same field. In the case of a simple rescaling of the $\vec{p}$ vector,
$\vec{\phi}=f(p)\vec{p}$, we perform explicitly the change of variable and define
$\vec{B}_e=f(p)\vec{X}_e$. This gives:
$$
\cZ^{(\phi)}_\Delta
\,=\,
\int \f{d^3B_e}{f(p_e)^3} \int dg_\es \; e^{i\sum_e \vec{B}_e\cdot\vec{p}_e}.
$$
If we now insert in the path integral $\cZ^{(\phi)}$ a $X_e^2$ for some edge $e$, it will thus
correspond to the insertion of a $B_e^2/f(p_e)^5$ term which obviously depends on the group
elements $g_\es$. We will compute this effect explicitly in the next section when studying the
length spectrum.

Let us point out that each choice of momentum vector $\vec{\phi}(G)$ defines a different Fourier
transform between $\SU(2)$ and $\R^3$ and induces a different covariant differential calculus on
the non-commutative $\R^3$ space \cite{karim}.

The issue behind this freedom in defining the discrete path integral can be seen from two
perspectives. Either we can see it as the problem of identifying which discrete $B_e$ variables are
truly the discretisation of the triad field $B$ in the continuum. Or we can consider it as an
ambiguity in the definition of the measure for the discrete path integral. We do not address this
issue in the present work.

\section{B-insertions in the Path Integral}
\label{bins}

We would like to discuss the insertion in the discretised BF path integral of amplitudes depending
on the $B$-field. Physically, this corresponds to either computing the average value of
$B$-observables or to taking into account a potential depending on the $B$-field. On the one hand,
$B$-observables represent geometric quantities since the triad $B$ defines the metric. On the other
hand, introducing a $B$-potential allows on to spinfoam quantise a whole class of ``constrained
topological theories" including Yang-Mills theories and gravity \cite{principle}.

Here, we focus on (ultra-)local $B$-insertions which depends on the $B$-variables through the norm
of the $\vec{B}_e$ variables without any coupling between the various edges of the triangulation. The
simplest case is the insertion of $B_e^2$ along some edge $e$. Mathematically, it means that
$\vec{B}_e$ is not anymore simply the Lagrange multiplier enforcing the triviality of the holonomy
around that edge. Physically, this calculation gives the ``length spectrum" for the spinfoam model.

\subsection{$B^2$-insertion and the length spectrum}

Let us focus on a single edge $e$ of the triangulation and insert $B_e^2$ in the path integral
along this edge. Instead of the original amplitude $\exp(i\vec{B}_e\cdot\vec{p}_e)$, we now have
the weight $B_e^2\exp(i\vec{B}_e\cdot\vec{p}_e)$. Integrating over $B_e$ does not give anymore the
$\tdelta(G_e)$ distribution. More precisely, we compute:
\be
\int \f{d^3\vec{B}}{4\pi}\,e^{i\vec{B}\cdot\vec{p}(G)}
\,=\,
2\sum_{j\in\N} d_j\chi_j(G),
\qquad
\int \f{d^3\vec{B}}{4\pi}\,B^2\,e^{i\vec{B}\cdot\vec{p}(G)}
\,=\,
2\sum_{j\in\N} [4j(j+1)-3]d_j\chi_j(G).
\ee
We extract the length spectrum from this formula:
\be
L_j^2= j(j+1)-\f34\,=\,\left(j+\f12\right)^2-1,\quad\forall j\in\N.
\ee
First, we point out that $L_0^2=-4/3<0$ for the trivial representation $j=0$ which defines a
negative ``vacuum area" and that $L_{1/2}=0$ would vanish for $j=1/2$. Second, this length spectrum
actually holds only for integer representations $j\in\N$. To obtain half-integer representations,
we need to work with $\SU(2)$. To this purpose, we insert the factor $(1+\f12\tr G)/2$. This gives
a slightly different and surprising formula:
\be
\f12\int \f{d^3\vec{B}}{4\pi}\,B^2\,\left(1+\f12 \tr G\right)\,e^{i\vec{B}\cdot\vec{p}(G)}
\,=\,
\sum_{j\in\N} [4j(j+1)-3]d_j\chi_j(G)
+\sum_{j\in\f12+\N} 4j(j+1)d_j\chi_j(G),
\ee
where the $-3$ shift of the length spectrum appears only for integer representations. Notice that
despite this shift, the length spectrum still increases with the spin $j$ without anomaly.

It is always possible to insert another factor $F(G)$ depending on the holonomy, thus changing the
measure on $G$, in order to shift arbitrarily the length spectrum~\footnotemark. There actually
exists an infinite number of measure factor reproducing the same length spectrum.

\footnotetext{
For instance, we consider an extra factor such as $\alpha+(1-\alpha)\chi_1(G)/3$ which is equal to
1 at $G=\pm\id$ and does not affect the partition function. Nevertheless, it changes the length
spectrum for $j\in\N$ leading to: $L_j^2=4j(j+1)-(1+8\alpha)/3$. For $\alpha=-1/8$, we use the
measure factor $(3\chi_1(G)-1)/8$ and obtain $L_j^2=4j(j+1)$. On the other hand, we can use the
measure factor $(\chi_1(G)-1)/2$ with $\alpha=-1/2$ and obtain simply $L_j^2=d_j^2$.}

In order to derive these formula, we have to compute the integral $\int d^3 B dG\,
\chi_j(G)B^2\exp(i\vec{B}\cdot\vec{p})$. There are various ways to deal with it. We can first
perform the integral over the group $\SU(2)$ and we obtain\footnotemark:
\be
\int dG\,\chi_j(G)\,e^{i\vec{B}\cdot\vec{p}}\,=\,\f{2}{|B|}J_{d_j}(|B|),
\ee
where the $J$'s are the Bessel functions.
\footnotetext{
We use the following spherical integral:
$$
\int_{\cS^2}\f{d^2\hat{u}}{4\pi}\,e^{i\vec{X}.\hat{u}}\,=\,
\f{\sin|X|}{|X|}.
$$}
This gives the probability distribution of the $B$-field and we get:
$$
\int d^3 B dG\, \chi_j(G)B^2e^{i\vec{B}\cdot\vec{p}}
\,=\,
\int_{\R_+} B^2dB\, \f{2J_{d_j}(B)}{B}.
$$
Unfortunately, this integral does not converge. Thus, in order to compute this integral, we use the
standard QFT method and introduce a source term $\exp(-i\vec{B}\cdot\vec{q})$. $B$-insertions will
be represented by differential operators of the type $\pp/\pp\vec{q}$ and we will send the source
$\vec{q}$ to 0 to obtain the final value. This gives for integer representations $j\in\N$:
\be
\f12\int \f{d^3 \vec{B}}{4\pi} dG\, \chi_j(G)B^2e^{i\vec{B}\cdot\vec{p}}
\,=\,
-\left(\f{\pp}{\pp q_k}\f{\pp}{\pp q_k}\right)\,
\left.\f{1}{\sqrt{1-q^2}}\chi_j(g)\right|_{q=0},
\ee
where the $1/\sqrt{1-q^2}$ factor comes from the Haar measure and the group element $g$ is given in
term of the vector source $\vec{q}$ as $g=\sqrt{1-q^2}+i\vec{q}.\vec{\sigma}$. This is a
straightforward Laplacian calculation in spherical coordinates:\footnotemark
$$
\left.\Delta\,\f{\chi_j(g)}{\sqrt{1-q^2}}\right|_0=
\left.\left(\pp_q^2+\f2q\pp_q\right)\f{\chi_j(g)}{\sqrt{1-q^2}}\right|_0
\,=\, -d_j^3+4d_j
\,=\,-d_j(d_j-2)(d_j+2).
$$
\footnotetext{
We can also notice that the function to differentiate is easily expressed in term of
$c\,\equiv\cos\theta=\sqrt{1-q^2}$:
$$
\f{\chi(g)}{\sqrt{1-q^2}}=\f{U_{2j}(c)}{c},
$$
in term of the Chebyshev polynomial (of the second kind). The evaluation of the Laplacian
$\Delta_q$ at $q=0$ actually collapses to a first order derivative in $c$:
$$
\left.\Delta_q\,f(c)\right|_{q=0}\,=\,-3f'(1),
$$
when $f'(1)$ is finite. Differentiating the Chebyshev polynomials can be easily done many different
ways. An elegant method is to use their generating functional:
$$
\sum_{n\in\N} U_n(\cos\theta)T^n\,=\,
\sum_{j\in\N/2}\chi_j(g)T^{2j}
\,=\,
\f{1}{1-2T\cos\theta+T^2}\,=\,
\f{1}{(T-e^{i\theta})(T-e^{-i\theta})}.
$$ }
%
%
Keeping in mind that $q=\sin\theta$, we compare this calculation  to the standard Casimir formula
when we act with the Laplacian on $\SU(2)$:
$$
\left.\Delta_\theta \chi_j(g)\right|_{g=\id} \,\equiv\,
\left.\f{1}{\sin^2\theta}\pp_\theta \sin^2\theta\pp_\theta\, \chi_j(g)\right|_{\theta=0}
\,=\,-4d_j (d_j-1)(d_j+1)
\,=\,-4d_j j(j+1).
$$

\subsection{Comparison with other path integrals}

It is natural to wonder how much does the length spectrum depend on the initial choice of path
integral. The effect is comparable to ordering ambiguities in the quantisation process. It will not
affect the leading order in $j(j+1)$ but will create a lower order shift compared to the previous
calculation. This entirely comes from the difference in the Haar measure factor in the various
parametrisation of the $\SU(2)$ group elements.

For instance, let us consider the path integral defined with the vector $\vec{u}=\theta\,\hat{u}$
and compute $\int d^3B dG\, B^2\chi_j(G)\exp(i\vec{B}\cdot\vec{u})$. Then the corresponding length
spectrum is given by the following Laplacian evaluation:
\be
\left.-\f{\pp}{\pp \vec{u}}.\f{\pp}{\pp \vec{u}}\,\f{\sin^2 u}{u^2}\chi_j(u)\right|_{u=0}.
\ee
Since the norm of the vector is simply $u=\theta$, this is equal to:
$$
-\left.\left(\pp_\theta^2+\f2\theta\pp_\theta\right)\,\f{\sin^2
\theta}{\theta^2}\chi_j(\theta)\right|_{\theta=0} =d_j(d_j^2+1).
$$
The difference between these results and previous length calculations \cite{principle,simone} in
the Ponzano-Regge model resides in the Haar measure factors.

This difference in the path integral measure affects the value of all observables depending on the
$B_e$ variables. There does not seem to be a unique choice of path integral and thus a unique
``length spectrum" in this formalism. We insist that this issue is not resolved by requiring that
the discrete path integral defines a topological state sum but it is an issue of the continuum
limit: which choice of discrete $\vec{B}_e$ variables represent faithfully the continuous triad
field $B$ (from which the metric is reconstructed).

\subsection{Polynomials $B$-insertions and further}

We can follow the same method as above to account for any polynomial insertion of the $B_e^{2n}$
type. Using the vector source and inserting more derivatives, we obtain:
\bes
\f12\int \f{d^3\vec{B}}{4\pi}dg \, B^2\chi_j(g)\,e^{\f12\tr Bg}&=&
(2j+3)(2j+1)(2j-1),\nonumber \\
\f12\int \f{d^3\vec{B}}{4\pi} dg\, B^4\chi_j(g)\,e^{\f12\tr Bg}&=&
(2j+5)(2j+3)(2j+1)(2j-1)(2j-3), \,\dots
\ees
and so on. This sequence as $n$ increases is actually related to the expansion of (modified) Bessel
functions. Indeed, let us now insert in the path integral a Gaussian weight $\exp(-\lambda B_e^2)$
i.e a quadratic potential in $B$. Such an insertion allows the spinfoam quantisation of 2d
Yang-Mills theory. We first perform the integration over $\vec{B}_e$:
$$
\int d^3\vec{B}\,e^{-\lambda B^2}e^{i\vec{B}.\vec{p}}
\,=\,
\left(\f{\pi}{\lambda}\right)^{\f32} e^{-\f{p^2}{4\lambda}}.
$$
Then, without even having to introduce a source, we compute (see \cite{graviton} for similar
calculations on $\SU(2)$ Gaussian states):
$$
\f12\int \f{d^3\vec{B}}{4\pi} dg\, e^{-\lambda B^2}\chi_j(g)\,e^{\f12\tr Bg}
=\f{1}{8(\pi\lambda)^{\f32}}
\int \f{d^3\vec{p}}{\sqrt{1-p^2}} \chi_j(g)e^{-\f{p^2}{4\lambda}}
\,=\,
\f{\sqrt{\pi}}{4\lambda^{\f32}}e^{-\f1{8\lambda}}\left[I_j\left(\f{1}{8\lambda}\right)-I_{j+1}\left(\f{1}{8\lambda}\right)\right].
$$
The previous polynomial insertions in $B^{2n}$ can then be extracted from an expansion of this
formula in $\lambda\sim 0$ (which corresponds to the asymptotic at infinity of the Bessel
functions).


\subsection{Feynman propagator insertion}

In order to take into account particles propagating on the spinfoam, we can also insert Feynman
propagators along the edges of the triangulated space-time manifold following the framework
introduced in \cite{PR3,effqg}. This allows one to add particles to the Ponzano-Regge model directly at
the path integral level.

We restrict ourselves to spinless particles with mass $m=\ka\sin\phi$ given in terms of the Planck
mass $\ka$ and an angle $\phi\in[0,\f\pi 2]$. Let us insert along a given edge $e$ the (undeformed)
Riemannian Feynman propagator, then the piece of the path integral concerning this edge reads:
\be
\int d^3\vec{B}_e\,
F_m(\vec{B}_e)\,e^{i\vec{B}_e\cdot\vec{p}_e},
\qquad
\textrm{with}\quad F_m(\vec{B})\,\equiv\,\f{e^{-i(\sin\phi-i\eps)|B|}}{4\pi|B|},\quad \eps\arr 0^+.
\ee
The integration over $\vec{B}_e$ is straightforward and gives the standard momentum representation
of the Feynman propagator as expected:
\be
\int d^3\vec{B}_e\, F_m(\vec{B}_e)\,e^{i\vec{B}_e\cdot\vec{p}_e}
\,=\, \f{1}{p_e^2-\sin^2\phi +i\eps}.
\ee
Following \cite{effqg}, this function on the $\SU(2)$ group is easily expandable in
representations:\footnotemark
\be\label{evenprop}
K_\phi(G)\,\equiv\,
\f{1}{p(G)^2-\sin^2\phi +i\eps}
\,=\,
\f{2}{\cos\phi}\sum_{j\in\N}e^{-id_j(\phi-i\eps)}\,\chi_j(G).
\ee
\footnotetext{
This formula can be checked directly since it only involves geometric series. One can otherwise
compute the coefficients of the decomposition into representations by evaluating the projection of
the Feynman propagator onto the $\SU(2)$ characters:
$$
\int dg\, \chi_j(g)\,\f{1}{|p(g)|^2-\sin^2 \phi+i\eps}
\,=\,
\f{2}{\pi}\int_0^\pi d\theta \,\f{\sin\theta \sin d_j\theta}{\sin^2\theta-\sin^2 \phi+i\eps}.
$$
This integral is evaluated using the residue formula for $j\ge 1$ and $\eps\arr 0^+$. It vanished
for half-integer spins $j\in\N+\f12$ and is equal to $2e^{-id_j\phi}/\cos\phi$ for integer spins
$j\in\N$. More precisely, we write the previous integral as a contour integral in the complex
plane:
$$
\f{1}{i\pi}\int_0^{2\pi}d\theta\f{\sin\theta e^{id_j\theta}}{\sin^2\theta-\sin^2\phi+i\eps}
=-\f{1}{\pi}\int_\cc dz\,z^{n-1}\f{X}{(X^2-\sin^2 \phi+i\eps)},
$$
where $X=\sin\theta=(z-z^{-1})/2i$ and $\cc$ is the unit circle in $\C$. The denominator can be
expanded and we find poles at $\pm\exp(\pm i\phi)$:
$$
\f{1}{(X^2-\sin^2 \phi+i\eps)}=
\f{(2iz)^2}{(z-e^{i\phi}-(1+i)\eps)(z+e^{-i\phi}-(1-i)\eps)(z-e^{-i\phi}+(1-i)\eps)(z+e^{i\phi}+(1+i)\eps)}.
$$}
This defines the corresponding Feynman propagator insertion in the spin foam amplitudes.  Moreover, it is given in
the usual format, in terms of $\SU(2)$ representations.

Such Feynman propagator insertions allow one to couple particles (and fields) to the spinfoam path
integral and to study the coupled and effective dynamics of both matter and gravitational sectors
\cite{PR3,effqg}.

\medskip

The insertion of an on-shell particle corresponds to the imaginary part of this propagator:
$$
\f{i}{4\tan\phi}\left[K_\phi(G)-K_{-\phi}(G)\right]=\sum_{j\in\N}\chi_j(\phi)\chi_j(G)=\tdelta_\phi(G),
$$
where $\tdelta(\cdot)$ is the $\SU(2)$ distribution identifying the class angle of $G$ to $\phi$ as
a $\SO(3)$ group element, that is, without distinguishing the angles $\phi$ and $(\pi-\phi)$. Also, the imaginary part of the initial $B$-insertion is:
$$
\f{i}{4\tan\phi}\left[F_m(|B|)-\overline{F_m}(|B|)\right]
\,=\,
\f{\sin(|B|\sin\phi)}{8\pi|B|\tan\phi}
\,=\,\f{\cos\phi}{32\pi^2}\int_{\cS^2}d^2\hat{u}\,e^{i\vec{B}.\sin\phi\,\hat{u}}.
$$
Written in this last form, the insertion of the massive on-shell particle is done directly at the
level of the discrete action (as it was first shown in \cite{PR1}, up to the $\cos\phi$ factor
which was discussed in \cite{PR3,oritla}).

\medskip

A last remark is about the $\SU(2)$ Feynman propagator instead of the $\SO(3)$
Feynman propagator used up to now. We define it, at the representation level, as the previous sum (\ref{evenprop})
extended to the whole of $\N/2$:
\be
\cos\phi\;\;k_\phi(G)\,\equiv\,2\sum_{j\in\N/2}e^{-id_j(\phi-i\eps)}\chi_j(G)
\,=\,
\f{1}{2\left(\sin^2\f\theta2-\sin^2\f\phi2+i\eps\right)}
\,=\,
\f{-1}{\cos\theta-\cos\phi - i\eps}.
\ee
We notice that summing over all half-integer spins cuts the angle by half and kills the $\SO(3)$
identification between the angles $\phi$ and $(\pi-\phi)$. We can not implement this
$\SU(2)$ Feynman propagator by a simple $B$-insertion. Indeed, any simple $B$-insertion in the
original path integral will naturally give solely $\SO(3)$ structures since it only depends on the
vector $\vec{p}(G)$. We need an insertion depending non-trivially on both $\vec{B}_e$ and the
holonomy $G_e$:
\be
k_{\phi_e}(G_e) = \int d^3\vec{B}_e\,
f_m(\vec{B}_e,G_e)\,e^{i\vec{B}_e\cdot\vec{p}_e},
\ee
with
\be
f_m(\vec{B},G)\,\equiv\,\left[1+\frac{1}{2\cos\phi}\chi_{\f12}(G)\right] \f{e^{-i(\sin\phi-i\eps)|B|}}{4\pi|B|},\; \eps\arr 0^+.
\ee

\section{B-observables and Gauge Fixing}
\label{observables}

Having discussed how the insertion of $B$-dependent functionals affects the discretised path
integral, we aim to evaluate the partition function with such $B$-insertions and to obtain the
value of such $B$-observables:
$$
\langle \cO \rangle = \frac{1}{\cN}\int \cD\omega\, \cD B\; \cO(B,\omega)\;e^{i\cS[B, \omega]},
$$
where the normalisation is given by $\cN = \cZ_\cM$. Both numerator and denominator need to be
gauge-fixed.

In the discrete setting, we choose a fixed edge $\bar{e}$ of the triangulated manifold $\Delta$ and we
seek to evaluate the average value of the length (squared) $B_{\bar{e}}^2$:
\bes
\cZ_\Delta &=& \prod_{e\in\Delta}\int d^3B_e \prod_{\es\in\Delta^*}\int dg_\es \; e^{i\cS[B_e,
g_\es]}, \\
\langle B_{\bar{e}}^2 \rangle&=& \f{1}{\cZ_\Delta} \prod_{e\in\Delta}\int d^3B_e \prod_{\es\in\Delta^*}\int dg_\es \;
B_{\bar{e}}^2\,e^{i\cS[B_e, g_\es]},
\ees
with the action $\cS=\sum_e \vec{B}_e\cdot\vec{p}(G_e)$.

First, this requires proper gauge fixing \cite{PR1}. We choose a maximal tree $T\in\Delta$ and we
fix $B_e=0$ for all edges $e\in T$. As we will show in an example later, we actually require that
$\bar{e}$ does not belong to $T$. Without such gauge fixing, we would always get a trivial (divergent)
result.  For this case, the Fadeev-Popov determinant is trivial, but as we show in Appendix \ref{fpdet}, for more complicated settings, this need not hold.

Second, we introduce a background ``state" $\psi(G_{\bar{e}})$, normalised such that $\psi(\id)=1$.
Inserting such an observable in the path integral does not affect the partition function, i.e
computing the value of $\langle \psi(G_{\bar{e}}) \rangle$ gives 1, since the path integral contains a
$\delta(G_{\bar{e}})$. On the other hand, such a state allows one to locate the edge $\bar{e}$: the state $\psi(G_{\bar{e}})$
excites different modes of the holonomy around the edge $\bar{e}$ and will modify the value of $B_{\bar{e}}^2$.
Thus we propose to compute:
\be
\langle B_{\bar{e}}^2\,\psi(G_{\bar{e}}) \rangle \equiv\f1{\cZ}\prod_{e\notin T}\int d^3B_e \prod_{\es\in\Delta^*}\int dg_\es
\; B_{\bar{e}}^2\psi(G_{\bar{e}})\,e^{i\cS[B_e, g_\es]},
\ee
for the choice $\psi(G_{\bar{e}})=\chi_j(G_{\bar{e}})/d_j$. The normalisation $d_j$ is to ensure that
$\psi(\id)=1$. The choice $j=0$ gives the vacuum value $\langle B_{\bar{e}}^2 \rangle$.

This setting ensures that computing $\langle B_{\bar{e}}^2\,\chi_j(G_{\bar{e}}) \rangle$ leads to the same
calculation as in the previous section. Indeed we find:
\bes
\f{1}{d_j}\,\langle B_{\bar{e}}^2\,\chi_j(G_{\bar{e}}) \rangle
&=& \f1{d_j}\int \f{d^3\vec{B}}{8\pi}dg \, B^2\chi_j(g)\,e^{\f12\tr Bg} \\
&=& (2j+3)(2j-1)
\,=\, 4j(j+1)-3
\,=\, 4\left(j+\f12\right)^2 -4. \nonumber
\ees
In particular, the vacuum value of $B_{\bar{e}}^2$ is surprisingly negative, $\langle B_{\bar{e}}^2 \rangle=-3$.
This can be compared to a non-zero vacuum energy. What matters for the spectrum is the length
variation:
\be
\delta L_j^2 \,\equiv\,
\f{1}{d_j}\langle B_{\bar{e}}^2 \chi_j(G_{\bar{e}}) \rangle - \langle B_{\bar{e}}^2\rangle
\,=\, 4j(j+1).
\ee
Thus, we recover the standard length (squared) spectrum given by the $\SU(2)$ Casimir operator.

\subsection{The tetrahedral triangulation}

\begin{figure}[h]
\begin{center}
\psfrag{a}{{\footnotesize 1}}
\psfrag{b}{\footnotesize 2}
\psfrag{c}{\footnotesize 3}
\psfrag{d}{\footnotesize 4}
\psfrag{1}{\tiny (12)}
\psfrag{2}{\tiny (13)}
\psfrag{3}{\tiny (14)}
\psfrag{4}{\tiny (23)}
\psfrag{5}{\tiny (24)}
\psfrag{6}{\tiny (34)}
\includegraphics[width = 6cm]{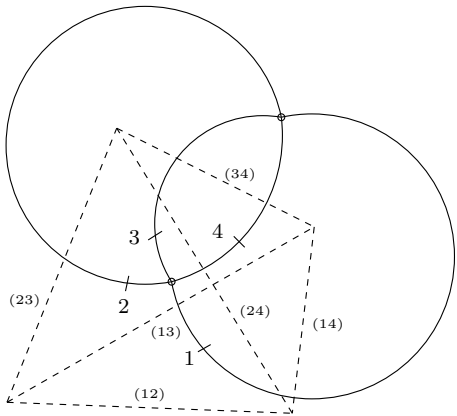}
\caption{\label{tetra} The triangulation of $S^3$ (dashed lines) and its dual (full lines). The triangles are labeled by single numbers, while the edges are coloured by pairs.  One can see immediately that the dual edges are in 1-1 correspondence with the triangles.  Furthermore, each dual face (plaquette) consists of two dual edges. }
\end{center}
\end{figure}

As an example to illustrate the gauge fixing procedure, we consider the triangulation of the
3-sphere $\cS^3$ with two tetrahedra.  This triangulation contains four vertices, the six edges connecting them, the four
triangles being these edges and the two 3-cells (inside and outside tetrahedra). The dual
triangulation (or spinfoam) has two dual vertices (representing the inside and outside tetrahedra)
and four dual edges connecting these two dual vertices and going through the four triangles.
Finally, there are six plaquettes (or dual faces) transverse to the six edges of the triangulation; see FIG. \ref{tetra}

We will number the four triangles (or dual edges) by $k=1,\dots,4$. We denote an edge by the (symmetric)
couple of triangles that share it $(kl)$. Then, the discretised partition function reads:
\be
\cZ\,=\,
\int \prod_{k<l} d^3\vec{B}_{kl} \int \prod_k dg_k\,
e^{\f1{2i}\sum_{k<l} \tr B_{kl}g_kg_l^{-1}}.
\ee
Let us consider the observable $\la B_{12}^2\ra$ defined as the quotient of the
$B_{12}^2$-insertion normalised by $\cZ$. In the case where we do not gauge fix the partition function, both
numerator and denominator contain divergent $\delta(\id)$ terms. Although these ``obvious" divergences cancel each other, we are left, nevertheless, with a meaningless (and divergent) correlation:
$$
\la B_{12}^2\ra \,=\, \f{\int d^3B \,B^2}{\int d^3B}.
$$
We now gauge fix the discrete path integral by setting all $\vec{B}_e$ variables to 0 along a
maximal tree in the triangulation. As  maximal tree, we choose $T=\{(23),(34),(41)\}$. We then
recover the result given above:
\be
\la B_{12}^2\ra\,=\,
\f{\int d^3B\int [dg]^{\times 4}\, B^2 e^{\tr Bg_1g_2^{-1}} \delta(g_1 g_3^{-1})\delta(g_2 g_4^{-1})}
{\int d^3B\int [dg]^{\times 4}\, e^{\tr Bg_1g_2^{-1}} \delta(g_1 g_3^{-1})\delta(g_2 g_4^{-1})},
\ee
where we have not imposed the $\delta(\cdot)$ constraints around the edges of the maximal tree.
After integrating over the superfluous group elements, we are left with:
\be
\la B_{12}^2\ra\,=\,
\f{\int d^3B\int dG \, B^2 e^{\tr (BG)}}
{\int d^3B\int dG \, e^{\tr (BG)}}\,=-3,
\ee
which we have evaluated previously. Computing $\la B_{12}^2\chi_j(g_1g_2^{-1})\ra$ does not change
the gauge fixing procedure.

The last subtlety is the case where we include the considered edge $(12)$ in the maximal
tree $T$. Then, the gauge fixing is not enough and we once again obtain a meaningless
result.

\subsection{The $\Theta$-graph and further $B$-observables}

In this final part, we generalise the previous $B_{\bar{e}}^2$ calculation to a coupled observable $\tr (B_a
B_b)$ probing the correlation between two edges, $a$ and $b$, of a triangle.\footnote{Were we to consider two arbitrary edges of the triangulation, we would have to rotate the $B$-fields into the same coordinate system, using the holomony variables, in order to arrive at a gauge invariant observable.} As a source for
the length, we use the following $\Theta$-state to excite the holonomies around the two edges
$a,b$:
\be
\psi(G_a,G_b)\,\equiv\, \f{1}{d_{j_a}d_{j_b}d_{J}}
\chi_{j_a}(G_a)\chi_{j_b}(G_b)\chi_J (G_a G_b).
\ee
We have normalised this state in order to ensure that $\psi(\id,\id)=1$, so that its insertion in
the partition function does not modify its evaluation. The representations $j_a,j_b$ respectively
excite the holonomies around the edges $a$ and $b$ while the representation $J$ couples the two
edges. Following the calculations of the previous section introducing vector sources
$\vec{q}_{a,b}$ for both edges, we find:
$$
\la \vec{B}_a\cdot\vec{B}_b\ra \,=\,
-\left.\f{\pp}{\pp\vec{q}_a}\f{\pp}{\pp\vec{q}_b}
\f{1}{\sqrt{(1-\vec{q}_a^2)(1-\vec{q}_b^2)}}\psi(g(\vec{q}_a),g(\vec{q}_b))\right|_{q=0}.
$$
This can be computed straightforwardly \footnotemark, as well as the average values
$\la\vec{B}_a^2\ra$ and $\la\vec{B}_b^2\ra$:
\bes
\la \vec{B}_a\cdot\vec{B}_b\ra &=& d_J^2-1, \\
\la\vec{B}_a^2\ra &=& (d_{j_a}^2-1) +(d_J^2-1)-3, \nonumber\\
\la\vec{B}_b^2\ra &=& (d_{j_b}^2-1) +(d_J^2-1)-3. \nonumber
\ees
When $J=0$, then $d_J^2=1$ and we recover the previous result for the length of a single edge.
\footnotetext{
The simplest method is to use the leading order behavior of the character:
$$
\chi_j(g(\vec{q}))\sim \,d_j\left(1+q^2\f{(1-d_j^2)}{6}\right).
$$}
We check that the scalar product $\la \vec{B}_a\cdot\vec{B}_b\ra$ is correctly smaller than the
product of the norms $\sqrt{\la\vec{B}_a^2\ra\la\vec{B}_b^2\ra}$ (as soon as $j_a\ge1$ or
$j_b\ge1$).

\section{Conclusion}
In this paper, we presented a short note on geometric observables in 3d discrete quantum gravity.
In particular, we concentrated on the accurate evaluation of polynomials in the $B$-field, within
the path integral approach.

The discretised version of the $BF$ theory path integral has the interesting property that is kills
half the degrees of freedom of the $\SU(2)$ gauge theory (the representations $j\in\N+\f12$).  It
maps it to an $\SO(3)$ gauge theory.  Fortunately, one can re-introduce them by inserting an
appropriate observable (\ref{fullpart}).   We noted that further ambiguity enters in the choice of
Lagrange multipliers ($B$ fields).


By far our main result was to compute the expectation value of gauge invariant polynomials of the
$B$ field.   We focused our attention on the length operator, and introduced the necessary gauge
fixing to measure the observable.  The gauge fixing occurred in two stages.  The first was to
introduce a normalised \lq\lq state'' which served to pick out the edge of interest and peak it on
a certain representation.   The second was the familiar gauge fixing of the $\isu(2)$ symmetry
using maximal trees.  We found that the expectation value of the length squared was actually
negative for the vacuum state ($j=0$) while it was greater than or equal to zero for all higher
representations.  A rather relevant fact is that the choice of discretisation of the $B$ field
affects the expectation value of the length variables.  Thus, while the ambiguity has no net effect
on the partition function, it has a very real consequence when observables are computed.

We undertook the generalisation of this process to more complicated observables, such as higher
order polynomials in the length, and other gauge invariant quantities, such as the dihedral angles
between two edges.  In particular, we could show that the edge vectors of a triangle satisfied the
Cauchy-Schwartz relation.

We elaborated on the topic of inserting an observable corresponding to the \lq\lq Feynman
propagator''  of a point particle.  Although, this concept is more an issue in the Lorentzian
theory \cite{lfdojr},  observables of an analogous functional form can be introduced in the
Riemannian regime.  We found that we can define such propagators  for both the $\SO(3)$ and
$\SU(2)$ theories.  Indeed, the \lq\lq Hadamard'' function, describing on-shell propagation, could
be defined in the usual fashion:  as the imaginary part of the Feynman propagator.

Finally, our work here has revealed an avenue to deal with $B$ observables in a more general context.   The next step would be to compute more coupled $B$-observables in order to study the discrete path integral for constrained $BF$ theories with a $B$-dependent potential such as Yang-Mills theory and gravity in more than three space-time dimensions.

\begin{appendix}
 \renewcommand{\theequation}{\thesection.\arabic{equation}}

\section{A few useful formulas}

A first interesting formula is the representation of the wave $\exp(i\vec{B}\cdot\vec{p})$ (see
e.g. \cite{russian1}):
\be
e^{\f12\tr (Bg)}\,=\,
\sum_{j\in\N/2} e^{-i\pi j}\,d_j\,\f{2J_{d_j}(|\vec{B}|)}{|\vec{B}|}\,\chi_j(e^{i\f\pi2\hat{B}\cdot\vec{\sigma}}g),
\ee
where we have decomposed the vector $\vec{B}=|B|\,\hat{B}$ into its norm and direction.

Secondly, we give the inverse formula allowing one to derive the characters from the wave and Bessel
insertions:
\be
\int d^3\vec{B}\, e^{i\vec{B}\cdot\vec{p}}\, \f{J_{d_j}(|\vec{B}|)}{4\pi|\vec{B}|}
\,=\,
\f1{|\vec{p}|} \int_0^{+\infty} dB\, \sin(B|\vec{p}|)J_{d_j}(B)
\,=\,
\f1{|\cos\theta|}\chi_j(\theta),
\ee
with $|\vec{p}|=\sin\theta>0$. The pre-factor is due to the Haar measure and is equal to
$\sqrt{1-p^2}=\eps\chi_{1/2}(g)$. This defines the Bessel $B$-observable that one has to insert in
the path integral in order to induce the character $\chi_j(G)$ on a plaquette instead of the basic
$\delta(G)$. This can be used when we want to ``gauge-fix" the  representation $j_e$ living on a
given edge $e$ to a fixed value.

\section{The Fadeev-Popov determinant for $B$-observables}
\label{fpdet}

We are interested in computing the Fadeev-Popov determinant in more detail. For $B$-observables, it emerges that it is non-trivial in many cases, although it was shown in \cite{lfdl} that the Fadeev-Popov determinant for pure gravity is trivial.

Let us consider the triangulation $\Delta$ and the graph $\Gamma$, upon whose edges one inserts arbitrary $B$-observables.  $\Gamma$ need not be connected but may consist of several components.  The important point is that the $B$-observables are $\SU(2)$ invariant but not translation invariant.  Thus, on a component $\Gamma_i\subset\Gamma$ with $|v|_i$ vertices, the translation symmetry is broken at $|v|_i-1$ of them.  This means that when one wishes to pick a maximal tree $T\subset \Delta$, it should only intersect each component  $\Gamma_i$ at one vertex exactly. Furthermore, when dealing with a tree, one can always pick a vertex to be the root, and orient the edges so that they point away from this root vertex. In particular, it should not contain an edge of $\Gamma$.  We shall neglect the gauge-fixing of the $\SU(2)$ symmetry as the observables are invariant and Fadeev-Popov determinant is trivial \cite{lfdl}.

Now that we have a maximal tree, we utilise the translation symmetry to set $B_e = 0$ for every $e\in T$.    We recall  that the discrete version of the translation symmetry (\ref{disctrans}) is:
$$
B_e \rar B_e + U_e^{t(e)}\phi_{t(e)} - [\Omega^{t(e)}_e, \phi_{t(e)}] -  U_e^{s(e)}\phi_{s(e)} + [\Omega^{s(e)}_e, \phi_{s(e)}],
$$
where we define the functions $U_e^v$ and $\Omega_e^v$ later on.

The inverse of the Fadeev-Popov determinant for such a symmetry is:
\be
D_{FP}^{-1} = \int\prod_{e\in T}  d\phi_e \;\delta\big(U_e^{t(e)}\phi_{t(e)} - [\Omega^{t(e)}_e, \phi_{t(e)}] -  U_e^{s(e)}\phi_{s(e)} + [\Omega^{s(e)}_e, \phi_{s(e)}]\big).
\ee
The Jacobian for the change of variables:
\be
\phi_e \rar \tilde{\phi}_e = U_e^{t(e)}\phi_{t(e)} - [\Omega^{t(e)}_e, \phi_{t(e)}] -  U_e^{s(e)}\phi_{s(e)} + [\Omega^{s(e)}_e, \phi_{s(e)}]
\ee
for every edge is:
\be
J\Big(\big\{\phi_e\big\}, \big\{\tilde{\phi}_e\big\}\Big) = \prod_{e\in T}\frac{1}{|U_e^{t(e)}|\left(|U_e^{t(e)}|^2 + 4|\vec{\Omega}_e^{t(e)}|^2\right)},
\ee
where $\Omega = i\vec{\Omega}\cdot \sigma$.
Thus, the Fadeev-Popov determinant is:
\be
D_{FP} = \prod_{e\in T} |U_e^{t(e)}|\left(|U_e^{t(e)}|^2 + 4|\vec{\Omega}_e^{t(e)}|^2\right).
\ee
Generically, the terms $U_e^v$ and $\Omega_e^v$ are functions of the holonomies associated to the edges incident at $v$ apart from the edge $e$ itself. They have the property that $U_e^v = 1$ and $\Omega_e^v = 0$ when the curvature of all the other incident edges vanishes.

This fact is enough to deal with the simplest case of a graph $\Gamma$ with only one connected component.  As mentioned above, the tree only hits $\Gamma$ at one vertex, and we shall call this vertex the root $v_{\text{root}}$.  For the moment, however, let us focus on a vertex at which the tree ends, that is, where only one edge of the tree is incident.  Then, the quantum amplitude ensures that the curvature around each of the edges $e\notin T$ incident at the vertex $v$ vanish.   Thus, the Fadeev-Popov factor for that edge is trivial.   Furthermore, the Bianchi identity ensures that the curvature associated edge $e\in T$ incident at $v$ also vanishes.  By applying that procedure to the outer edges and working our way towards the root of the tree, we can show that the whole Fadeev-Popov factor is trivial.

For the case of a graph $\Gamma$ with two components, we call the vertices where the tree $T$ intersects $\Gamma_1,\,\Gamma_2$ as $v_{\text{root}}$ and $v_{\text{term}}$ respectively.   In particular, there is a unique path $\mathfrak{p}\subset T$ joining $v_{\text{root}}$ and $v_{\text{term}}$.  When we attempt to apply the same procedure as above, we find that the Fadeev-Popov factors are trivial for edges $e\in T/ \mathfrak{p}$. But for edges $e\in\mathfrak{p}$, we find that the curvature associated to the two edges $e\in\mathfrak{p}$ incident at a vertex  are non-vanishing.  Thus, the Fadeev-Popov factor for these edges factor non-zero.

We shall calculate $U_e^v$ and $\Omega_e^v$ explicitly in the case $v\in\mathfrak{p}$.  We know that the translation symmetry is satisfied thanks to the Bianchi identity:
\be
\prod_{e @ v} g_e = \mathbb{I},
\ee
where $g_e$ is the holonomy associated to the edge $e$.  But the curvature vanishes for all but the two edges $e\in\mathfrak{p}$, so the identity reduces to:
\be
g_1g_2 = (u_1\mathbb{I} +P_1)(u_2\mathbb{I}+P_2) = \left(u_1u_2 +\frac{1}{2}\tr(P_1P_2)\right)\mathbb{I} + \left(u_1P_2 + u_2P_1 +\frac{1}{2}[P_1,P_2]\right) = \mathbb{I}.
\ee
Then, the transformation of the relevant B variables is:
\bes
B_1&\rar& B_1+ u_2\phi +\frac{1}{4}[P_2,\phi],\\
B_2 &\rar& B_2 + u_1\phi - \frac{1}{4}[P_1,\phi].
\ees
Say that $e_1$ terminates at $v$.  The Fadeev-Popov determinant for that edge is:
\be
|U_{1}^{v}|\left(|U_1^{v}|^2 + 4|\vec{\Omega}_1^{v}|^2\right) = |u_2|\left(|u_2|^2+\frac{1}{4}|\vec{p}_2|^2\right) = |\cos\theta_2|\left(\cos^2\theta_2+\frac{1}{4}\sin^2\theta_2\right).
\ee
But the Bianchi identity imposes that $\theta_1=-\theta_2 =: \theta$.  Therefore, the Fadeev-Popov determinant for this tree finishes up as:
\be
D_{FP} = \left[|\cos\theta|\left(\cos^2\theta+\frac{1}{4}\sin^2\theta\right)\right]^{|\mathfrak{p}|}.
\ee
where we have assumed that the tree is chosen such that $v_{\text{term}}$ lies as an endpoint of
$\Gamma_2$.  Should it lie in the interior, the Fadeev-Popov factor for the edge $e\in T$ incident
at $v_{\text{term}}$ will be more complicated since more than two edges at $v_{\text{term}}$ will
have non-vanishing curvature.

For the more general case of graphs with multiple components, the Fadeev-Popov factors are trivial
on all the edges apart from those on the unique path in $T$ which join the vertices in
$T\cap\Gamma$.    Also, apart from the case where $\Gamma$ has just one component,  the
Fadeev-Popov determinant, and hence the resulting amplitude, depends on the choice of tree.

\end{appendix}


\end{document}